\documentclass{article}
\usepackage{spconf}
\usepackage{amsmath}
\usepackage{graphicx}
\usepackage{cite}
\usepackage{mathptmx}  
\usepackage{multirow}
\usepackage{array}
\usepackage{booktabs}
\usepackage{subcaption}
\usepackage{colortbl}
\usepackage{amssymb}
\usepackage[utf8]{inputenc}
\usepackage[normalem]{ulem}
\usepackage{comment}
\usepackage{balance}
\usepackage{siunitx}
\usepackage{hyperref}
\usepackage{url}
\usepackage{pifont}
\usepackage{setspace} 
\usepackage{soul}

\newcommand{\cmark}{\ding{51}}
\newcommand{\xmark}{\ding{55}}

\title{OSegNet: Operational Segmentation Network for COVID-19 Detection using Chest X-ray Images}

\name{Aysen Degerli$^{\dagger}$\thanks{This study was supported by the NSF-Business Finland Center for Visual
and Decision Informatics (CVDI) Advanced Machine Learning for Industrial
Applications (AMaLIA) project under Grant 4183/31/2021.}, Serkan Kiranyaz$^{\ast}$, Muhammad E. H. Chowdhury$^{\ast}$, and Moncef Gabbouj$^{\dagger}$}

\address{$^{\dagger}$Faculty of Information Technology and Communication Sciences, Tampere University, Tampere, Finland \\
$^{\ast}$Department of Electrical Engineering, Qatar University, Doha, Qatar}

\begin{document}
\ninept
\maketitle

\begin{abstract}
Coronavirus disease 2019 (COVID-19) has been diagnosed automatically using Machine Learning algorithms over chest X-ray (CXR) images. However, most of the earlier studies used Deep Learning models over scarce datasets bearing the risk of overfitting. Additionally, previous studies have revealed the fact that deep networks are not reliable for classification since their decisions may originate from irrelevant areas on the CXRs. Therefore, in this study, we propose Operational Segmentation Network (OSegNet) that performs detection by segmenting COVID-19 pneumonia for a reliable diagnosis. To address the data scarcity encountered in training and especially in evaluation, this study extends the largest COVID-19 CXR dataset: QaTa-COV19 with $121,378$ CXRs including $9258$ COVID-19 samples with their corresponding ground-truth segmentation masks that are publicly shared with the research community. Consequently, OSegNet has achieved a detection performance with the highest accuracy of $99.65\%$ among the state-of-the-art deep models with $98.09\%$ precision. 
\end{abstract}

\begin{keywords}
SARS-CoV-2, COVID-19, Machine Learning, Deep Learning
\end{keywords}

\section{Introduction}
Coronavirus disease 2019 (COVID-19), caused by severe acute respiratory syndrome coronavirus-2 (SARS-CoV-2), has infected millions after it was first reported in 2019. The World Health Organization (WHO) has declared COVID-19 as a pandemic since it is highly contagious (especially its mutations), and affects seriously immunocompromised patients and elderly \cite{vishnevetsky2020rethinking}. However, performing a reliable diagnosis of COVID-19 is challenging since it reveals similar symptoms such as cough, breathlessness, and fever compared to other viral diseases  \cite{singhal2020review}. Moreover, COVID-19 may not be always symptomatic, causing asymptomatic individuals to spread the disease to population \cite{10.3389/fpubh.2020.00473}. Consequently, computer-aided diagnosis is necessary to perform fast and accurate COVID-19 detection to prevent the further spread of the disease.

COVID-19 diagnosis can be performed via nucleic acid detection with real-time polymerase chain reaction (RT-PCR) and imaging techniques: computed tomography (CT) and chest X-ray (CXR) imaging. Even though RT-PCR is defined as the reference standard to diagnose COVID-19, it lacks stability in the laboratory test results with high false-negatives rate \cite{tahamtan2020real}. Contrary to RT-PCR, CT has higher sensitivity level \cite{bernheim2020chest}. However, its clinical utility is limited especially for asymptomatic individuals \cite{waller2020diagnostic}. Thus, CXR imaging is widely used due to its fast acquisition, easy accessibility, less radiation exposure, and lower risk of cross-infection among other diagnostic tools \cite{cozzi2020chest}.

Deep Learning (DL) has achieved a remarkable performance in the COVID-19 diagnosis using CXRs. Many studies \cite{narin2021automatic, apostolopoulos2020covid, wang2020covid, chowdhury2020pdcovidnet, pham2020classification} used DL models to perform COVID-19 classification by transfer learning. However, they have evaluated the performance of deep networks only over scarce and limited size datasets. The data scarcity has the potential to cause overfitting since DL models need significantly large amount of data for generalization. Moreover, the control group of the aforementioned studies contains only healthy subjects or limited thoracic diseases, i.e. bacterial or other viral pneumonia against COVID-19 pneumonia. Thus, their clinical usage is unfeasible for real-case scenarios. Additionally, several studies \cite{degerliICIP, keidar2021covid, tahir2022deep} have investigated the decision-making process of deep models in classification tasks. Accordingly, the unreliability of DL models was revealed by the activation maps, where their attention was on the irrelevant areas of CXRs, such as background, text, or bones rather than the lungs. Thus, few studies \cite{degerli2021covid, tahir2021covid} performed COVID-19 pneumonia segmentation for a reliable COVID-19 detection with deep networks using CXRs.

\begin{figure*}[t!]
    \centering
    \includegraphics[width=.99\linewidth]{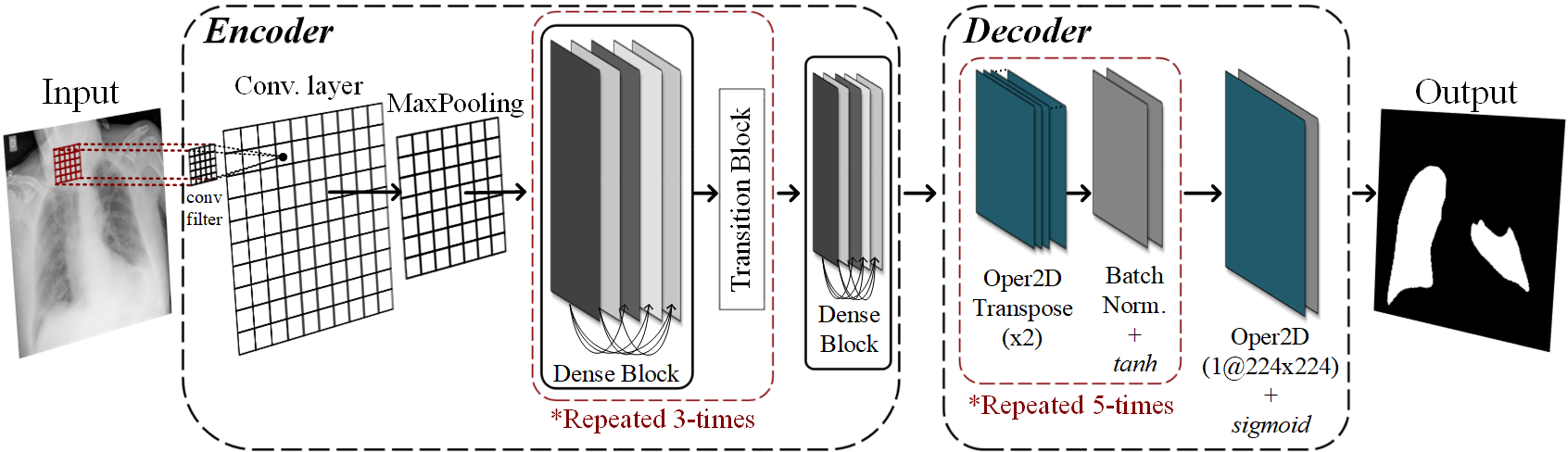}
    \caption{The proposed OSegNet model for COVID-19 pneumonia segmentation is illustrated, where the transfer learning is performed at the encoder block, and operational layers (Oper2D) are used at the decoder block.}
    \label{fig:segmentation}
\end{figure*}

In this study, to address the aforementioned limitations, we propose \textbf{O}perational \textbf{Seg}mentation \textbf{Net}work (OSegNet) that performs COVID-19 pneumonia segmentation for the diagnosis using CXR images. Contrary to convolutional layers used in many deep networks, operational layers with generative neurons of Self-Organized Operational Neural Networks (Self-ONNs) \cite{KIRANYAZ2021294, malik2021self, yilmaz2021self, kelecs2021self, devecioglu2021real} are used in the decoder block. Self-ONNs are heterogeneous network models with generative neurons that can create any non-linear transformation in each kernel element. Such diversity does not only yield a superior learning performance but also allows a significant reduction in the network depth and complexity. The proposed OSegNet has an autoencoder structure except that operational layers are used at the decoder as illustrated in Fig. \ref{fig:segmentation}. Thus, this study uses operational layers for the first time for image segmentation. Additionally, in this study, the QaTaCOV-19 dataset that was introduced previously by our study \cite{degerli2021covid} is extended to reach $9258$ COVID-19 samples with their corresponding ground-truth segmentation masks. Thus, together with a control group of $112,120$ CXRs from healthy subjects and $14$ different thoracic diseases, QaTa-COV19\footnote{The benchmark QaTa-COV19 is publicly shared at the repository \href{https://www.kaggle.com/aysendegerli/qatacov19-dataset}{https://www.kaggle.com/aysendegerli/qatacov19-dataset}.} is the largest publicly available dataset for COVID-19 pneumonia segmentation over CXR images.

The rest of the paper is organized as follows. The proposed OSegNet model and QaTaCOV-19 dataset are introduced in Section \ref{sec:methods}. The experimental results and conclusion are given in Section \ref{sec:experiments} and Section \ref{sec:conclusion}, respectively.

\section{Methodology and Materials}\label{sec:methods}
In this section, the proposed OSegNet model is first introduced, and then, the details of the benchmark QaTa-COV19 dataset are presented.

\subsection{OSegNet: Operational Segmentation Network}
Convolutional Neural Networks (CNNs) are widely used for many computer vision tasks including COVID-19 diagnosis. However, the potential of CNNs is limited due to the homogeneous network structure and linear neuron model. Thus, many studies have proposed deeper structures with skipping connections to diverse the modality of CNNs for boosting their performance. Furthermore, the performance is increased with transfer learning that helps to faster and stable convergence of the model.

Contrary to convolutional layers, operational layers \cite{KIRANYAZ2021294} have a generative neuron model that can create any non-linear transformation of each kernel element to achieve a highly heterogeneous network. Accordingly, each neuron input, $x_l$ at layer, $l$ is calculated as follows:
\begin{equation}
    x_l = b_l + \sum_{j=1}^{N}\Phi(w_l^j, y_{(l-1)}^j),
\end{equation}
where $b$ is the bias, $N$ is the number of neurons at the previous layer, and a nodal operation, $\Phi$ is performed between the weights of the layer, $w_l$ and the outputs of the previous layer, $y_{(l-1)}$. Nodal operator functions are generated during back-propagation training using Taylor polynomial approximation of any non-linear function. Thus, nodal operator functions can define any arbitrary function, $f(x)$ as the infinite sums of the function's derivatives at a point $(x=a)$ as follows:
\begin{equation}
    f(x) = \sum_{n=0}^{\infty}\frac{f^{(n)}(a)}{n!}(x-a)^n, 
\end{equation}
where $f^{(n)}(a)$ is the $n^{\text{th}}$ derivative of $f$ at the point $a$, and $n!$ is the factorial of $n$. Accordingly, nodal operator functions can be truncated by the $Q^{\text{th}}$ order Taylor approximation as follows:
\begin{equation}
    \Phi(\mathbf{w}, y) = \sum_{q=0}^Q w_q (y-a)^q,
    \label{eq3}
\end{equation}
where $\mathbf{w}$ is the array that contains weights  $w_q=\frac{f^{(q)}(a)}{q!}$. The Maclaurin series representation of (\ref{eq3}) can be formulated for $a=0$ using the tangent hyperbolic (\textit{tanh}) activation function that maps the neuron outputs into $[-1 ,1]$ as follows:
\begin{equation}
    \Phi(\mathbf{w}, y) = \sum_{q=1}^Q w_q (y)^q,
\end{equation}
where $w_0$ is dropped due to the compensation from the common bias element, $b$ of each neuron. The structure of OSegNet is similar to an autoencoder that maps the input image, $\mathbf{I}$ to its output mask, $\mathbf{M}:\mathbf{M}\leftarrow{\Upsilon_{\varepsilon, \delta}(\mathbf{I})}$, where the network $\Upsilon$ consists of encoder $\epsilon$, and decoder $\delta$ parts as depicted in Fig. \ref{fig:segmentation}. Accordingly, the OSegNet encoder is composed of a state-of-the-art model, where its weights are initialized with the ImageNet weights by transfer learning. The proposed model has operational layers as decoding the features of the state-of-the-art model, where the decoder $\delta\in\{b_l, w_l\}_{l=1}^L$ consists of $L$ number of operational layers composed of five decoder blocks. Each decoder block includes an operational transpose layer for upsampling by $\times2$, batch normalization, and \textit{tanh} activation function. The output of the last block is attached to an operational layer with \textit{sigmoid} activation function. For each operational layer, kernel size of $k=(3\times3)$ is used sequentially with the filter sizes of $\{128, 64, 32, 16, 8, 1\}$. Finally, OSegNet is trained over $S$ number of samples $\{\iota_{train}^i, \mu_{train}^i\}_{i=1}^{S}$, where $\iota$ and $\mu$ are training data and ground-truth masks, respectively.

\vspace{0.1cm}

In this study, the state-of-the-art networks: DenseNet-121 \cite{huang2017densely} and Inception-v3 \cite{szegedy2016rethinking}, where their weights are initialized with the ImageNet weights by transfer learning are used as the encoder of the OSegNet model. Additionally, the decoder structures: UNet++ \cite{zhou2018unet++} and DLA \cite{yu2018deep} that merges encoder and decoder with skipping connections and nested convolutional blocks used as the competing networks against the proposed model. 

\begin{table*}[t!]
\centering
\caption{COVID-19 \textit{pneumonia segmentation} performance results (\%) computed over the test (unseen data) set of QaTa-COV19 dataset using state-of-the-art and the proposed OSegNet models.}
\vspace{-0.2cm}
\begin{tabular}{|c|c|cccccc|}
\hline
\textbf{Encoder} & \textbf{Model} & \textbf{Sensitivity} & \textbf{Specificity} & \textbf{Precision}  & \textbf{F1-Score} & \textbf{F2-Score}  & \textbf{Accuracy} \\ \hline \hline
 
\multirow{7}{*}{DenseNet-121} & UNet++ & $83.16$ & $\textbf{99.91}$ & $89.56$  & $86.24$ & $84.37$  & $99.76$ \\

& DLA & $84.65$ & $\textbf{99.91}$ & $89.13$  & $86.83$ & $85.51$ & $99.77$\\

& OSegNet $(Q=1)$ & $83.56$ & $\textbf{99.91}$ & $89.60$  & $86.47$ & $84.70$ & $99.76$  \\

& OSegNet $(Q=2)$ & $86.32$ & $99.89$ & $87.57$ & $86.94$ & $86.56$ & $99.76$ \\

& OSegNet $(Q=3)$ & $\textbf{87.25}$ & $99.89$ & $87.58$  & $\textbf{87.42}$ & $\textbf{87.32}$  & $99.77$ \\

& OSegNet $(Q=4)$ & $84.96$ & $\textbf{99.91}$ & $\textbf{89.85}$ & $87.33$ & $85.89$ & $\textbf{99.78}$  \\

&  OSegNet $(Q=5)$ & $86.88$ & $99.86$ & $85.23$  & $86.04$ & $86.54$ & $99.74$  \\ \hline \hline

\multirow{7}{*}{Inception-v3} & UNet++ & $88.95$  & $99.86$ & $85.33$  & $87.10$ & $88.20$ & $99.76$ \\

& DLA & $86.23$  & $\textbf{99.91}$ & $\textbf{89.63}$ & $\textbf{87.89}$ & $86.89$ & $\textbf{99.78}$  \\ 
 
& OSegNet $(Q=1)$ & $86.47$ & $99.78$ & $78.18$ & $82.12$ & $84.67$ & $99.66$\\ 

& OSegNet $(Q=2)$ & $88.09$ & $99.88$ & $87.31$ & $87.70$ & $87.93$ & $\textbf{99.78}$ \\

& OSegNet $(Q=3)$ & $\textbf{89.36}$ & $99.87$ & $86.37$  & $87.84$ & $\textbf{88.75}$ & $\textbf{99.78}$ \\ 

& OSegNet $(Q=4)$ & $87.70$ & $99.89$ & $87.69$ & $87.70$ & $87.70$ & $\textbf{99.78}$ \\
 
& OSegNet $(Q=5)$ & $88.33$ & $99.88$ & $86.78$  & $87.55$ & $88.02$ & $99.77$\\  \hline

\end{tabular}
\label{tab:segmentation-results}
\end{table*} 

\subsection{QaTa-COV19 Dataset}
Tampere University and Qatar University researchers have compiled the QaTa-COV19 dataset that is the largest CXR dataset for COVID-19 pneumonia segmentation. The control group images of the dataset are obtained from ChestX-ray14 dataset \cite{wang2017chestx} that consists of $112,120$ CXRs from healthy subjects and $14$ different thoracic diseases. Additionally, $9258$ COVID-19 images are collected from the publicly available BIMCV-COVID19+ dataset \cite{vaya2020bimcv} along with the CXRs from our previous study \cite{degerli2021covid}. In this study, we annotated the CXRs of BIMCV-COVID19+ \cite{vaya2020bimcv} to create the extended version of QaTa-COV19. For this purpose, we have first eliminated the acquisitions from the same patient, session, and run in BIMCV-COVID19+ \cite{vaya2020bimcv} to remove any duplications.

\begin{table}[t!]
\centering
\caption{Details of QaTa-COV19 dataset.}
\resizebox{.48\textwidth}{!}{
\begin{tabular}{|c|c|c|c|c|}
\hline
\rowcolor[gray]{.90}Data & \begin{tabular}[c]{@{}c@{}}Training \\ Samples\end{tabular} & \begin{tabular}[c]{@{}c@{}}Augmented\end{tabular} & \begin{tabular}[c]{@{}c@{}}Augmented \\ Training Samples\end{tabular} & \begin{tabular}[c]{@{}c@{}}Test \\ Samples\end{tabular} \\ \hline \hline

\begin{tabular}[c]{@{}c@{}}ChestX-ray14\end{tabular} & $86,524$ & \xmark & $86,524$ & $25,596$ \\ \hline

\begin{tabular}[c]{@{}c@{}}COVID-19 \\ \end{tabular} & $7145$ & \cmark & $20,000$ & $2113$ \\  \hline \hline

Total & 93,669 &  & $\textbf{106,524}$ & $\textbf{27,709}$ \\ \hline
\end{tabular}}
\label{tab:numberofsamples}
\end{table}

The ground-truths of CXRs are generated by the collaborative human-machine annotation approach that enables fast and accurate annotation of COVID-19 pneumonia regions using deep networks inspired by U-Net \cite{ronneberger2015u}, UNet++ \cite{zhou2018unet++}, and DLA\cite{yu2018deep} architectures as used in our previous study \cite{degerli2021covid}. These networks are trained by previously annotated $2951$ COVID-19 samples and $12,544$ healthy subjects that are from the group-I data in \cite{degerli2021covid}. The trained segmentation networks are used to predict the \textit{ground-truth masks} of $6307$ CXRs from BIMCV-COVID19+ \cite{vaya2020bimcv}. Accordingly, the best predictions of the segmentation networks are selected as the ground-truth segmentation masks by the collaboration of expert medical doctors. At last, the predicted segmentation masks of only $31$ CXR images are not selected since they are not accurate enough; hence, they are manually drawn by medical doctors.

Table \ref{tab:numberofsamples} shows the details of QaTa-COV19 dataset. Since the train and test sets of ChestX-ray14 \cite{wang2017chestx} are predefined, COVID-19 samples are split with the same train/test ratio as in \cite{wang2017chestx} by taking the patient information into account; thus, they contain different subjects. The CXRs in QaTaCOV-19 dataset are resized to $224\times224$ pixels. We have applied data augmentation using the Image Data Generator in Keras. Accordingly, CXRs are randomly rotated in a $10-$degree range and $10\%$ shifted vertically and horizontally with the \textit{nearest} mode to fill pixels outside the input boundaries. 

\section{Experimental Evaluation}\label{sec:experiments}
In this section, the experimental setup is introduced. Then, the experimental results are reported over the QaTa-COV19 dataset. 

\subsection{Experimental Setup}

The experimental evaluations are performed over the test (unseen) set of the QaTaCOV-19 dataset. COVID-19 pneumonia segmentation is evaluated on a pixel level, where foreground (pneumonia) and background are considered as positive-class and negative-class, respectively. Accordingly, the standard performance metrics are calculated as follows: \textit{sensitivity} is the ratio of correctly identified COVID-19 samples in the positive class, \textit{specificity} is the rate of correctly detected control group samples in the negative class, \textit{precision} is the ratio of correctly detected COVID-19 samples among the samples that are detected as positive class, \textit{accuracy} is the ratio of correctly identified samples in the dataset. Lastly, the \textit{F-score} is defined as follows: 
\begin{equation}
    F(\beta)-Score = (1+\beta^2)\frac{(\textit{precision}\times\textit{sensitivity})}{\beta^2\times\textit{precision}+\textit{sensitivity}},
\end{equation}
where \textit{F$1-$Score} is the harmonic average between \textit{sensitivity} and \textit{precision} for $\beta=1$, whereas \textit{F$2-$Score} tolerates \textit{sensitivity} metric for $\beta=2$. Accordingly, the objective is to achieve a high \textit{sensitivity} level and \textit{F$1-$Score} as minimizing the false alarm ($1-\textit{specificity}$).

The networks are implemented with the TensorFlow library on NVidia ® GeForce RTX 2080 Ti GPU card. For the optimizer, we have used Adam with its default parameter settings. Furthermore, a hybrid loss function is used that combines dice and focal loss by summation. Let the ground-truth mask be $\mathbf{K}$, where the pixel label is $\kappa$, and the model prediction is $\hat{\kappa}$. Accordingly, the probabilities are defined as $P(\kappa=1)=p$ and $P(\hat{\kappa}=1)=q$. Thus, we define the dice loss as follows:
\begin{equation}
    D(p, q) = 1 - \frac{2 \sum p_{h, \omega} q_{h, \omega}}{\sum p_{h, \omega} + \sum q_{h, \omega}},
\end{equation}
where $h$ and $\omega$ are the indices of height and width of the CXRs. Furthermore, the focal loss is defined as follows:
\begin{equation}
    F(p, q) = -\alpha (1-q)^\gamma p \log q - (1 - \alpha) q^\gamma (1-p)\log(1-q),
\end{equation}
where the parameters are set as $\gamma=2$ and $\alpha=0.25$. Accordingly, models are trained over $50-$epochs with a learning rate of $10^{-4}$.

\begin{table*}[t!]
\centering
\caption{COVID-19 \textit{detection} performance results (\%) computed over the test (unseen data) set of QaTa-COV19 dataset using state-of-the-art and the proposed OSegNet models.}
\vspace{-0.2cm}
\begin{tabular}{|c|c|cccccc|}
\hline
\textbf{Encoder} & \textbf{Model} & \textbf{Sensitivity} & \textbf{Specificity} & \textbf{Precision} & \textbf{F1-Score} & \textbf{F2-Score} & \textbf{Accuracy} \\ \hline \hline
 
\multirow{7}{*}{DenseNet-121} & UNet++ & $94.32$ & $\textbf{99.87}$ & $\textbf{98.37}$ & $96.30$ & $95.10$ & $99.45$  \\

& DLA & $94.60$ & $99.76$ & $97.04$ & $95.81$ & $95.08$ & $99.37$ \\

& OSegNet $(Q=1)$ & $\textbf{98.15}$ & $99.65$ & $95.89$ & $97.01$ & $\textbf{97.69}$ & $99.54$  \\

& OSegNet $(Q=2)$ & $97.68$ & $99.71$ & $96.58$ & $97.13$ & $97.46$ & $99.56$  \\

& OSegNet $(Q=3)$ & $97.59$ & $99.70$ & $96.45$ & $97.01$ & $97.36$ & $99.54$ \\

& OSegNet $(Q=4)$ & $97.44$ & $99.83$ & $97.95$ & $\textbf{97.70}$ & $97.55$ & $\textbf{99.65}$ \\

& OSegNet $(Q=5)$ & $95.55$ & $99.78$ & $97.30$ & $96.42$ & $95.90$ & $99.46$  \\ \hline \hline

\multirow{7}{*}{Inception-v3} & UNet++ &  $\textbf{98.53}$  & $99.56$ & $94.85$  & $96.66$ & $\textbf{97.77}$ & $99.48$ \\

& DLA & $96.78$  & $\textbf{99.84}$ & $98.08$ & $97.43$ & $97.04$ & $99.61$  \\
 
& OSegNet $(Q=1)$ & $97.87$ & $97.93$ & $79.57$ & $87.78$ & $93.57$ & $97.92$ \\ 

& OSegNet $(Q=2)$ & $97.54$ & $99.78$ & $97.35$ & $97.45$ & $97.50$ & $99.61$  \\

& OSegNet $(Q=3)$ & $97.35$ & $\textbf{99.84}$ & $\textbf{98.09}$ & $\textbf{97.72}$ & $97.50$ & $\textbf{99.65}$ \\ 

& OSegNet $(Q=4)$ & $97.35$ & $99.82$ & $97.81$ & $97.58$ & $97.44$ & $99.63$ \\
 
& OSegNet $(Q=5)$ & $98.01$ & $99.73$ & $96.78$ & $97.39$ & $97.76$ & $99.60$ \\  \hline

\end{tabular}
\label{tab:detection-results}
\end{table*}

\subsection{Experimental Results}
In this section, we report the performances of COVID-19 pneumonia segmentation and detection. The COVID-19 pneumonia segmentation results are shown in Table \ref{tab:segmentation-results}, where state-of-the-art and the proposed OSegNet models are compared. The variation in the performance of OSegNet is investigated by changing the $Q$ parameter. Primarily, we have observed that each model has achieved a successful pneumonia segmentation with an F$1-$Score of $>86\%$ and specificity of $>99.75\%$. It can be seen from Table \ref{tab:segmentation-results} that any model with the encoder of Inception-v3 outperforms DenseNet-121 simply due to its complex structure and higher number of trainable parameters. Accordingly, among state-of-the-art models, 
the best segmentation performance has been achieved by the duo of UNet++ and Inception-v3 with an F$2-$Score of $88.20\%$. Nevertheless, the OSegNet $(Q=3)$ with Inception-v3 encoder has achieved the highest sensitivity level of $89.36\%$ and F$2-$Score of $88.75\%$ among all. 

\begin{table}[b!]
\centering
\caption{Confusion matrices of the best computing UNet++ and the proposed OSegNet $(Q=3)$ models with Inception-v3 encoders for COVID-19 detection.}
\vspace{-0.2cm}
\begin{subtable}{.45\textwidth}
\centering
\vspace{-0.1cm}
\caption{Confusion Matrix UNet++}
\begin{tabular}{|c|c|c|c|}
\hline
\multicolumn{2}{|c|}{\multirow{2}{*}{\textbf{UNet++}}} & \multicolumn{2}{c|}{Predicted} \\ \cline{3-4} 
\multicolumn{2}{|c|}{} & \multicolumn{1}{c|}{Control Group} & \multicolumn{1}{c|}{COVID-19} \\ \hline
\multirow{2}{*}{\begin{tabular}[c]{@{}c@{}}Ground\\ Truth\end{tabular}} & Control Group & $25483$ & $113$ \\ \cline{2-4} 
 & COVID-19 & $31$ & $2082$ \\ \hline
\end{tabular}
\end{subtable}

\bigskip
\noindent
\begin{subtable}{.45\textwidth}
\centering
\vspace{-0.2cm}
\caption{Confusion Matrix OSegNet $(Q=3)$}
\begin{tabular}{|c|c|c|c|}
\hline
\multicolumn{2}{|c|}{\multirow{2}{*}{\textbf{OSegNet}}} & \multicolumn{2}{c|}{Predicted} \\ \cline{3-4} 
\multicolumn{2}{|c|}{} & \multicolumn{1}{c|}{Control Group} & \multicolumn{1}{c|}{COVID-19} \\ \hline
\multirow{2}{*}{\begin{tabular}[c]{@{}c@{}}Ground\\ Truth\end{tabular}} & Control Group & $25556$ & $40$ \\ \cline{2-4} 
 & COVID-19 & $56$ & $2057$ \\ \hline
\end{tabular}
\end{subtable}
\label{tab:CMs}
\end{table}

The detection performances are presented in Table \ref{tab:detection-results}, which are calculated per CXR sample. Accordingly, a CXR sample is classified as COVID-19 if \textit{any} pixel in the output mask is predicted as COVID-19 pneumonia. The duo of UNet++ and Inception-v3 holds the best detection performance among state-of-the-art with the highest sensitivity level of $98.53\%$. Nevertheless, the highest F$1-$Score of $97.72\%$ and accuracy of $99.65\%$ has been achieved once again by the OSegNet $(Q=3)$ model. Accordingly, the confusion matrices of the best computing models: UNet++ and OSegNet $(Q=3)$ with Inception-v3 encoders are given in Table \ref{tab:CMs}. It is observed that UNet++ only misses $31$ COVID-19 cases, whereas OSegNet $(Q=3)$ has lower false alarms with only $40$ samples. Lastly, Table \ref{table-parameters} shows that OSegNet $(Q=3)$ model has $3.6$M and $1.2$M less number of parameters with faster inference time compared to the UNet++ model for both DenseNet-121 and Inception-v3 versions, respectively.

\begin{table}[b!]
\centering
\caption{The number of trainable and non-trainable parameters of the models with their inference time (ms) per sample.}
\vspace{-0.3cm}
\resizebox{.48\textwidth}{!}{
\begin{tabular}{c|c|ccc|}
\cline{2-5}
 & \multicolumn{1}{c|}{\textbf{Model}} & \multicolumn{1}{c}{\textbf{Trainable}} & \multicolumn{1}{c}{\textbf{Non-Trainable}} & \multicolumn{1}{c|}{\textbf{Time}} \\ \hline
 
\multicolumn{1}{|c|}{\multirow{7}{*}{{\rotatebox[origin=c]{90}{DenseNet-121}}}} & UNet++ & $14.40$M & $88.45$K & $4.9115$ \\

\multicolumn{1}{|c|}{} & DLA & $13.15$M  & $88.45$K & $4.4265$ \\

\multicolumn{1}{|c|}{} & OSegNet $(Q=1)$ & $8.23$M & $84.14$K & $1.8652$ \\

\multicolumn{1}{|c|}{} & OSegNet $(Q=2)$ & $9.51$M & $84.14$K & $2.0525$ \\ 

\multicolumn{1}{|c|}{} & OSegNet $(Q=3)$  & $10.79$M & $84.14$K & $2.2201$ \\ 

\multicolumn{1}{|c|}{} & OSegNet $(Q=4)$ & $12.07$M & $84.14$K & $2.3932$ \\

\multicolumn{1}{|c|}{} & OSegNet $(Q=5)$ & $13.34$M & $84.14$K & $2.5674$
\\ \hline \hline

\multicolumn{1}{|c|}{\multirow{7}{*}{{\rotatebox[origin=c]{90}{Inception-v3}}}} & UNet++   & $30.43$M & $39.23$K & $5.0247$\\

\multicolumn{1}{|c|}{} & DLA & $28.96$M & $39.23$K & $4.5111$\\

\multicolumn{1}{|c|}{} & OSegNet $(Q=1)$ & $24.23$M & $34.93$K & $1.7533$ \\

\multicolumn{1}{|c|}{} & OSegNet $(Q=2)$ & $26.68$M & $34.93$K & $2.0046$\\

\multicolumn{1}{|c|}{} & OSegNet $(Q=3)$ & $29.14$M & $34.93$K & $2.1807$ \\

\multicolumn{1}{|c|}{} & OSegNet $(Q=4)$ &  $31.60$M & $34.93$K & $2.3575$ \\

\multicolumn{1}{|c|}{} & OSegNet $(Q=5)$ &  $34.06$M & $34.93$K & $2.5469$ \\

\hline 

\end{tabular}}
\label{table-parameters}
\end{table}  

\section{Conclusions}\label{sec:conclusion}
\vspace{-0.1cm}
Computer-aided diagnosis plays a vital role in the COVID-19 detection to prevent the further spread of the disease. As a major contribution, this study publicly shares the largest CXR dataset, QaTa-COV19 which consists of $9258$ COVID-19 samples with their corresponding ground-truth segmentation masks along with $112,120$ control group CXRs. The experimental results over the QaTa-COV19 dataset show that the proposed OSegNet model has achieved the highest sensitivity level of $89.36\%$ for the COVID-19 segmentation, and precision of $98.09\%$ for the COVID-19 detection while the network complexity and depth has been reduced.

\begin{spacing}{.9}
\bibliographystyle{IEEEbib}
\bibliography{refs}
\end{spacing}

\end{document}